\documentclass[prl,showpacs,twocolumn,floatfix]{revtex4}
\usepackage{graphicx}
\usepackage{bm}
\usepackage{hyperref}
\def\he#1{$\rm^{#1}He $}

\begin{document}

\title{The effective mass of two--dimensional $^3$He }
\author{J. Boronat$^\dagger$, J. Casulleras$^\dagger$, V. Grau$^\dagger$,
E. Krotscheck$^\ddagger$ and J. Springer$^\ddagger$}
\affiliation{Departament de Fisica i Enginyera Nuclear, Campus Nord B4-B5,
Universitat Polit\'ecnia de Catalunya\\
E-08034 Barcelona, Spain\\}
\affiliation{$^\ddagger$Institut f\"ur Theoretische Physik, Johannes
Kepler Universit\"at, A 4040 Linz, Austria}

\begin{abstract}
We use structural information from diffusion Monte Carlo calculations
for two--dimensional \he3 to calculate the effective mass. Static
effective interactions are constructed from the density-- and spin
structure functions using sumrules. We find that both spin-- and
density-- fluctuations contribute about equally to the effective
mass. Our results show, in agreement with recent experiments, a flattening of
the single--particle self--energy with increasing density, which
eventually leads to a divergent effective mass.
\end{abstract}

\maketitle

Two--dimensional liquid \he3 is particularly interesting because it
is, even at zero temperature, not self--bound and can, therefore, be
studied in a wide density range. Although governed by one of the
simplest Hamiltonians for realistic many--body systems, \he3 exhibits
a wide range of delicate and complex phenomena which have,
by--and--large, been resilient to a understanding from the underlying
Hamiltonian. Only recently, Monte Carlo techniques have moved to a
point where structural properties have been understood from first
principles \cite{BoronatHe3,Boronat2DHe3}.

Low--energy dynamical properties of \he3 at low temperatures are
phenomenologically described by Landau's Fermi--Liquid theory, which
establishes relationships between observable quantities such as the
specific heat, the compressibility, and the magnetic susceptibility.
Understanding the so--called Fermi--Liquid parameters in \he3 has
therefore been a recurring issue in theoretical low--temperature
research. The calculation of Fermi--Liquid parameters in terms of
Feynman diagrams is operationally well defined, but the execution of
the theory from an underlying microscopic Hamiltonian is far too
complicated to be practical. Hence, many attempts have been made to
explain the features of Fermi--Liquid parameters within
semi--phenomenological models \cite{ber66,don66,vol84}.

We examine in this paper physical effects contributing to the
effective mass in two--dimensional \he3. This work is motivated by a
recent sequence of measurements \cite{Saunders2dmass} that seem to
indicate a Mott--Hubbard transition in quasi--two--dimensional \he3
atomic monolayers.  Technically, our calculations correspond to those
of Ref. \onlinecite{Bengt,he3mass}, but we will use as much
information as possible from accurate ground state Monte Carlo simulations.

\begin{figure}[ht]
        \centerline{\includegraphics[width=0.8\linewidth]{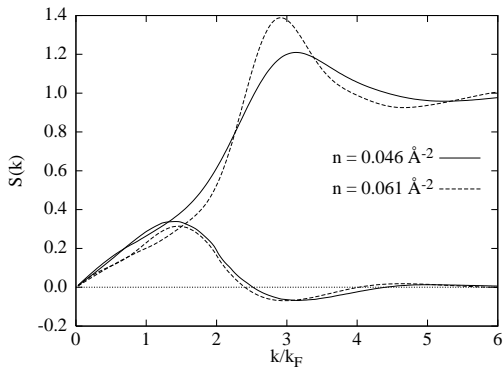}}
\caption{The density structure functions $S(k)$ (upper curves)
and correlation part of the spin structure functions
$S_\sigma(k)-S_F(k)$ (lower curves) are shown at the densities $\rho =
$~0.046~\AA$^{-2}$ (solid lines) and $\rho = $~0.061~\AA$^{-2}$
(dashed line). The zero level (dotted line) is included as a guide to
the eye.
\label{fig:SofkPieces}}
\end{figure}

The relevant quantity for the effective mass is the single--particle
propagator $G(k,\omega)$ in the vicinity of the Fermi surface.  It is
expressed in terms of the proper self--energy $\Sigma^*(k,\omega)$
through the Dyson equation \cite{FetterWalecka}
\begin{equation}
G_{\sigma\sigma'}
(k,\omega) = {\delta_{\sigma\sigma'}
 \over \hbar\omega - t(k) - \Sigma^*(k,\omega)}\,;
\label{eq:Dyson}
\end{equation}
$t(k) = \hbar^2 k^2/2m$ is the free single--particle spectrum.  The
physical excitation spectrum is obtained by finding the poles of the
Green's function in the $(k,\omega)$--plane. Several steps are
involved in constructing practically useful expressions for the proper
self--energy $\Sigma^*(k,\omega)$.  The first step is the derivation
of effective interactions. We use for that purpose results from
diffusion Monte--Carlo calculations \cite{Boronat2DHe3}. The structure
function can be written as
\begin{equation}
S(k) = S_\rho(k) + S_\sigma(k){\bm \sigma}_1\cdot{\bm \sigma}_2 \,,
\label{eq:Ssigma}
\end{equation}
where the components are constructed from the structure
functions for parallel and antiparallel spins. Both quantities are
obtained by either directly evaluating the expectation value of
$\rho_{\bf k}\rho_{-{\bf k}}$ or by Fourier transforming the
corresponding pair distribution functions. The first procedure is more
accurate for long wavelength up to the size of the simulation box,
whereas the latter is appropriate for medium and short wave
lengths. Moreover, one can determine the long--wavelength limit of
$S_\rho(k)$ from the bulk compressibility; we will comment on this
below. We have taken the Fourier transform of the pair distribution
function for wave numbers $k\ge k_F$, and direct simulation data for
long wavelength, and have smoothly interpolated these data towards
$k\rightarrow 0$. The static structure functions at two densities are
shown in Fig. \ref{fig:SofkPieces}. The density static structure
function shows the typical behavior, whereas the spin--structure
function depends only weakly on the density in the most interesting
regime.

The static structure functions are related to the dynamic response
functions through the $m_0$ sumrule
\begin{equation}
S_s(k) = - \int_0^\infty {d(\hbar\omega)\over\pi} \Im m \chi_s(k,\omega)\,.
\label{eq:Ssum0}
\end{equation}
Above, $s \in \{\rho,\sigma\}$ refers to the spin channel.  Assuming a
model like the random phase approximation (RPA),
\begin{equation}
\chi_s(k,\omega) =
{\chi_0(k,\omega)\over 1- \tilde V_s(k) \chi_0(k,\omega)}\,,
\label{eq:chiRPA}
\end{equation}
we can relate the static structure functions $S_s(k)$ uniquely to the
effective interactions $\tilde V_s(k)$. The tilde in potential
indicates that we use dimensionless Fourier transforms,
$\tilde V_s(k) = \rho\int d^3r V_s(r) e^{i{\bf k}\cdot{\bf r}}$,
correspondingly the Lindhard function has the dimension of an inverse
energy. We note in passing that the
random phase approximation (\ref{eq:chiRPA}) also satisfies the $m_1$
sumrule as an identity, whereas the inclusion of at least
two--particle--two--hole excitations is needed to satisfy
higher--order sumrules \cite{bosegas}.

The long--wavelength limit of $\tilde V_\rho(k)$ is related to the
bulk compressibility
\begin{equation}
{d\over d\rho}\rho^2 {d E\over d \rho}
= {d\over d\rho}\rho^2 {d E_F\over d \rho} + \tilde V_\rho(0+)\,,
\end{equation}
where $E$ and $E_F$ are the energy per particle of the interacting
and the noninteracting systems, respectively.
We can thus
determine $\tilde V_\rho(0+)$ from the equation of state.

\begin{figure}[ht]
\vspace{0.2truein}
	\centerline{\includegraphics[width=0.8\linewidth]{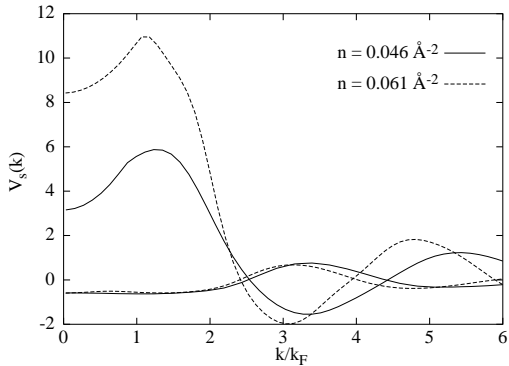}}
\caption{The density--channel and spin--channel
interactions obtained from the corresponding structure functions
through the RPA relationship (\ref{eq:Ssum0}) are shown for the
densities $\rho = $~0.046~\AA$^{-2}$ (solid lines)
and $\rho = $~0.061~\AA$^{-2}$ (dashed lines).
The interactions are given in units of $\hbar^2 k_F^2/2m$.
\label{fig:veffs}}
\end{figure}

Fig. \ref{fig:veffs} shows the effective interactions $\tilde V_s(q)$
defined through the relations (\ref{eq:Ssum0}), (\ref{eq:chiRPA}) at
two representative densities. These effective interactions resemble
the Aldrich--Pines pseudopotentials \cite{Aldrich} which have been
derived in a similar spirit. The most prominent features are the same:
the density--channel interaction is repulsive and can lead to a zero
sound excitation, whereas the spin--channel interaction does not. We
also note that, similar to the spin--structure function, there is
relatively little change in the spin--channel effective interaction
when given in dimensionless units.

Getting back to the self--energy, we assume low--lying excitations. In
that case, the so--called G0W approximation
\cite{FetterWalecka,FrimanBlaizot} for the self--energy,
\begin{widetext}
\begin{equation}
\Sigma(k,E) = u(k) + i\sum_s(2 s + 1)\displaystyle{\int
{ d^2 q \, d(\hbar\omega)\over
\rho (2\pi)^3}} G^0({\bf k}-{\bf q},E-\hbar\omega) \tilde
V_s^2(q)\chi_s(q,\omega)\label{eq:G0W}
\equiv u(k) + \Sigma^{(\rho)}(k,E) +  \Sigma^{(\sigma)}(k,E)
\end{equation}
\end{widetext}
should be appropriate. We have split the full self--energy into an
energy--independent mean field term $u(k)$ and the two dynamic, energy
dependent portions $\Sigma^{(\rho)}(k,E)$ and $\Sigma^{(\sigma)}(k,E)$
originating from coupling to density-- and spin--fluctuations,
respectively.  We have taken for $u(k)$ the exchange term of the
static density--channel interaction $\tilde V_\rho(q)$. One could here
in principle also use the single--particle spectrum of correlated
basis functions theory, but the basic results are very similar and
within the limits of the present description.

The self--energy is conveniently evaluated by Wick rotation in the
complex $\omega$- plane, the salient features have been discussed in
the literature \cite{Bengt,FrimanBlaizot,he3mass}.  With the stated
approximations, one obtains the spectrum
\begin{equation}
\epsilon(k) = t(k) + \Sigma(k,\epsilon(k))\,.
\label{eq:eofk}
\end{equation}
In the numerical applications, we have used the ``on-shell
approximation'' $\epsilon(k)\rightarrow t(k)$ in the self--energy. It
is our experience in \he3-\he4 mixtures that this gives good agreement
with much more sophisticated implementations of the same theory
\cite{mixmass}. Especially, one might be led to ``dress'' the
single--particle Green's functions in the self--energy (\ref{eq:G0W})
by solving Eq. (\ref{eq:eofk}). However,
single--particle Green's functions appear in two locations: One is the
external propagator spelled out explicitly in Eq. (\ref{eq:G0W}), the
other location is the particle--hole propagator.  To maintain the
symmetry between ``internal'' and ``external'' propagators, one should
apply any modifications to either both, or to none. Using a
non--trivial spectrum the particle--hole propagator is, on the other
hand, extremely dangerous because one would then violate the $m_0$ and
$m_1$ sumrules and, hence, modify the overall importance of the
dynamic part of the self--energy in an uncontrolled way. A {\it
systematic\/} improvement of the theoretical framework is the
inclusion of {\it pair\/} excitations \cite{KarSLT23}.

\begin{figure}[ht]
	\centerline{\includegraphics[width=0.8\linewidth]{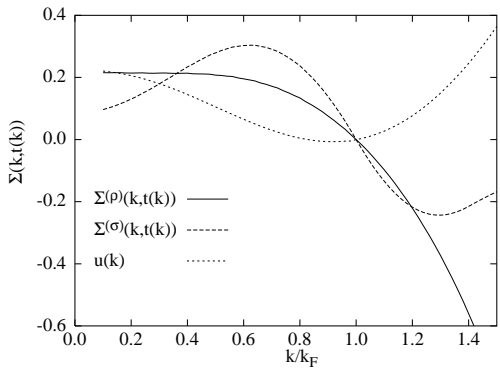}}
\caption{The figure shows the Fock term $u(k)$ (short--dashed line),
the ``density'' term $\Sigma_\rho(k,t(k))$ and the
the ``spin'' term $\Sigma_\sigma(k,t(k))$ (long--dashed line),
and their sum (solid line) of the density channel self-energy
for the density  $\rho = $~0.046~\AA$^{-2}$.
All functions have been shifted to be zero at the Fermi momentum,
all energies are given in units of $\hbar^2 k_F^2/2m$.
\label{fig:SigmaPieces}}
\end{figure}

\begin{figure}[ht]
	\centerline{\includegraphics[width=0.8\linewidth]{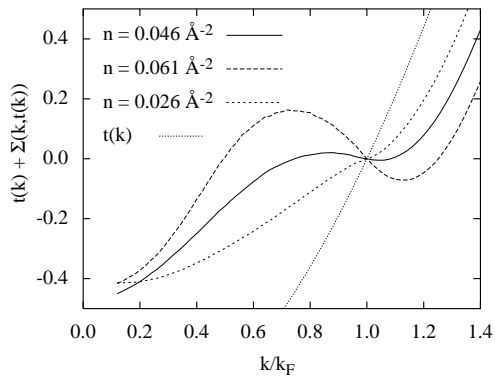}}
\caption{The figure shows the full on--shell spectrum
(\protect\ref{eq:eofk}) for the densities $\rho = $~0.026~\AA$^{-2}$,
$\rho = $~0.046~\AA$^{-2}$, and $\rho = $~0.061~\AA$^{-2}$.  The free
single--particle spectrum $t(k)$ is also shown for comparison.  All
functions have been shifted to be zero at the Fermi momentum, all
energies are given in units of the Fermi--energy of the
non--interacting liquid.
\label{fig:SigmaDens}}
\end{figure}

Fig. \ref{fig:SigmaPieces} shows the mean field $u(k)$ and the dynamic
contributions $\Sigma^{(\rho)}(k,t(k))$ and
$\Sigma^{(\sigma)}(k,t(k))$. All three terms are, in the vicinity of
the Fermi--momentum, rather smooth functions of the single--particle
momentum $k$. $u(k)$ has positive slope at the Fermi wave number and,
hence, decreases the effective mass whereas both dynamic contributions
cause an effective mass enhancement. The $S$-shape of the
spin--channel term is typical for attractive interactions
\cite{ZaringhalamMass,PethickMass,Bengt,he3mass}. At higher momenta,
we found similar structures as in three--dimensional \he3 originating
from a coupling of the single--particle excitation to the maxon
\cite{Bengt,he3mass}; these will not be discussed here. The full
on--shell self--energy is shown, for several densities, in
Fig. \ref{fig:SigmaDens}. The most evident feature is the development
of a saddle around the Fermi wave number, which is due to
spin--fluctuations and leads, at high densities, to an
instability.

\begin{figure}[ht]
        \centerline{\includegraphics[width=0.8\linewidth]{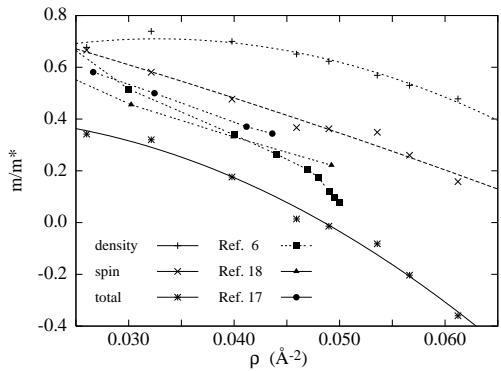}}
\caption{The figure shows the density--dependence of $m/m^*$ at the
Fermi wave number $k_F$ (so lid line). Also shown are the results from
density--fluctuations (long--dashed line) and from spin--fluctuations
only (short--dashed line). The dashed lines with filled markers show
the experimental values of Refs. \protect\onlinecite{Greywall90},
\protect\onlinecite{Godfrin2dmass}, and
\protect\onlinecite{Saunders2dmass}. The lines through the theoretical
data are quadratic fits.
\label{fig:MofRho}}
\end{figure}

The effective mass is obtained from the
single--particle spectrum through
\begin{equation}
{\hbar^2 k_F\over m^*} \equiv \left.{d \epsilon(k)\over dk}\right|_{k=k_F}\,.
\label{eq:mofk}
\end{equation}
Both, experiments \cite{Greywall90,Godfrin2dmass,Saunders2dmass} and
our calculations, indicate that the effective mass increases rapidly
with density and eventually becomes singular. The primary quantity
that one calculates is the single particle spectrum. For a theoretical
analysis, it is therefore more convenient to discuss the inverse,
$m/m^*$.  Fig. \ref{fig:MofRho} shows, as our final result, the
density dependence of the effective mass ratio $m/m^*$ and compares it
with the experiments of Refs. \onlinecite{Greywall90},
\onlinecite{Godfrin2dmass}, and \onlinecite{Saunders2dmass}.  The
fluctuations of our results are due to the statistical uncertainties
of the Monte Carlo simulations. These are most pronounced in the
spin--dependent correlations since the spin--structure function,
$S_\sigma(k) = S_{\uparrow\downarrow}(k) - S_{\uparrow\uparrow}(k)$,
is the difference of two quantities obtained from DMC simulations, but
these fluctuations do not affect our general result.

The overall theoretical picture is practically the same as the
experimental one, although our theory overestimates the correlation
effect somewhat and the instability occurs around 0.048~\AA$^{-2}$.
The experimental ratio $m/m^*$ is, to reasonable accuracy, a linear
function of the density which goes through zero between $\rho =
0.051\,$\AA$^{-2}$ and $\rho = 0.07\,$\AA$^{-2}$, causing a
singular $m^*$. Our theoretical calculations reproduce the
experimental values within about 10-20 percent when compared with the
spectrum of the non--interacting Fermi gas. This is satisfactory
considering the simplicity of the G0W approximation (\ref{eq:G0W}).

The individual contributions from spin-- and density fluctuations are
also shown in Fig. \ref{fig:MofRho}. The mean field term and density
fluctuations contribute somewhat less than spin--fluctuations,
but are non--negligible. Both effects are individually insufficient to
reproduce the experimental values. This is the case in both two and
three dimensions \cite{he3mass}, but the relative importance of
spin--fluctuations appears to be larger in 2D.

In both, three and two dimensions, one observes a divergence of the
effective mass at some high density. Extrapolating the data of
Ref. \onlinecite{GRE83,GRE86}, one observes that the divergence of the
effective mass would appear in 3D at a density 0.03 \AA$^{-3}$,
whereas the liquid--solid phase transition occurs at 0.023 \AA$^{-3}$.
The 2D situation is somewhat different: Ref.
\onlinecite{Saunders2dmass} implies a divergence of $m^*$ at about
0.051~\AA$^{-2}$; earlier data \cite{Greywall90,Godfrin2dmass} suggest
a somewhat higher density. The film freezes between 0.052~\AA$^{-2}$
\cite{Saunders2dmass} and 0.063~\AA$^{-2}$ \cite{GodfrinULT98}.

It seems unlikely that freezing and the singularity of the effective
mass have the same cause. The divergence in $m^*$ is due to the
increasing importance of spin fluctuations with density. This is
manifested very clearly in 2D and also visible in 3D. The theory used
here reproduces those features, in both 3D and 2D, at a
semi--quantitative level without the need for phenomenological input.
The fact that we obtain a negative slope of the single particle
spectrum is clearly a consequence of the G0W approximation; a
self--consistent theory should not have solutions for unstable
situations. Nevertheless, relatively simple approximations have often
shown the same physics as more sophisticated theories in the stable
regime, who then simply cease to have solutions beyond the point of an
instability.  We are presently not prepared to speculate on ``what's
beyond'' the singularity.

A second interesting question is why the 2D theory apparently
overestimates the effective mass, whereas it underestimates $m^*$ in
3D. One can only speculate that a 2D model is an obvious
simplification of the real physical situation of an adsorbed film,
and little is known about the severity of such an approximation for
\he3.

We have shown that understanding the value of the effective
mass in two--dimensional \he3 is -- in analogy to the more common
three--dimensional case -- not a simple problem, and simple paradigms
that try to attribute the effect to a single cause are genuinely
inadequate. Both spin-- and density--fluctuations have profound
effects, although spin--fluctuations are stronger in 2D and we are
more inclined to associate the effect of density fluctuations to
Feynman-Cohen backflow instead of ``localization''.  We hesitate to
identify the flattening of the single--particle spectrum with a Mott
transition, it may well indicate either a transition to ``anomalous
occupation numbers'' or indicate simply a lack of self--consistency of
the calculation of the spectrum. Quantitative improvement will first
be sought in a more accurate description of the response functions
\cite{KarSLT23}.

\section*{ACKNOWLEDGMENTS}

This work was supported, in part, by the Austrian Science Fund under
project P12832-TPH and from DGI (Spain) Grant No. BFM2002-00466 and
Generalitat de Catalunya Grant No. 2001SGR-00222.

\begin{thebibliography}{21}
\expandafter\ifx\csname natexlab\endcsname\relax\def\natexlab#1{#1}\fi
\expandafter\ifx\csname bibnamefont\endcsname\relax
  \def\bibnamefont#1{#1}\fi
\expandafter\ifx\csname bibfnamefont\endcsname\relax
  \def\bibfnamefont#1{#1}\fi
\expandafter\ifx\csname citenamefont\endcsname\relax
  \def\citenamefont#1{#1}\fi
\expandafter\ifx\csname url\endcsname\relax
  \def\url#1{\texttt{#1}}\fi
\expandafter\ifx\csname urlprefix\endcsname\relax\def\urlprefix{URL }\fi
\providecommand{\bibinfo}[2]{#2}
\providecommand{\eprint}[2][]{\url{#2}}

\bibitem[{\citenamefont{Casulleras and Boronat}(2000)}]{BoronatHe3}
\bibinfo{author}{\bibfnamefont{J.}~\bibnamefont{Casulleras}} \bibnamefont{and}
  \bibinfo{author}{\bibfnamefont{J.}~\bibnamefont{Boronat}},
  \bibinfo{journal}{Phys. Rev. Lett.} \textbf{\bibinfo{volume}{84}},
  \bibinfo{pages}{3121} (\bibinfo{year}{2000}).

\bibitem[{\citenamefont{Grau et~al.}(2002)\citenamefont{Grau, Boronat, and
  Casulleras}}]{Boronat2DHe3}
\bibinfo{author}{\bibfnamefont{V.}~\bibnamefont{Grau}},
  \bibinfo{author}{\bibfnamefont{J.}~\bibnamefont{Boronat}}, \bibnamefont{and}
  \bibinfo{author}{\bibfnamefont{J.}~\bibnamefont{Casulleras}},
  \bibinfo{journal}{Phys. Rev. Lett.} \textbf{\bibinfo{volume}{89}},
  \bibinfo{pages}{045301} (\bibinfo{year}{2002}).

\bibitem[{\citenamefont{Berk and Schrieffer}(1966)}]{ber66}
\bibinfo{author}{\bibfnamefont{N.~F.} \bibnamefont{Berk}} \bibnamefont{and}
  \bibinfo{author}{\bibfnamefont{J.~R.} \bibnamefont{Schrieffer}},
  \bibinfo{journal}{Phys. Rev. Lett.} \textbf{\bibinfo{volume}{17}},
  \bibinfo{pages}{433} (\bibinfo{year}{1966}).

\bibitem[{\citenamefont{Doniach and Engelsberg}(1966)}]{don66}
\bibinfo{author}{\bibfnamefont{S.}~\bibnamefont{Doniach}} \bibnamefont{and}
  \bibinfo{author}{\bibfnamefont{S.}~\bibnamefont{Engelsberg}},
  \bibinfo{journal}{Phys. Rev. Lett.} \textbf{\bibinfo{volume}{17}},
  \bibinfo{pages}{750} (\bibinfo{year}{1966}).

\bibitem[{\citenamefont{Vollhardt}(1984)}]{vol84}
\bibinfo{author}{\bibfnamefont{D.}~\bibnamefont{Vollhardt}},
  \bibinfo{journal}{Rev. Mod. Phys.} \textbf{\bibinfo{volume}{56}},
  \bibinfo{pages}{99} (\bibinfo{year}{1984}).

\bibitem[{\citenamefont{Casey et~al.}(2003)\citenamefont{Casey, Patel,
  Ny{\'e}ki, Cowan, and Saunders}}]{Saunders2dmass}
\bibinfo{author}{\bibfnamefont{A.}~\bibnamefont{Casey}},
  \bibinfo{author}{\bibfnamefont{H.}~\bibnamefont{Patel}},
  \bibinfo{author}{\bibfnamefont{J.}~\bibnamefont{Ny{\'e}ki}},
  \bibinfo{author}{\bibfnamefont{B.~P.} \bibnamefont{Cowan}}, \bibnamefont{and}
  \bibinfo{author}{\bibfnamefont{J.}~\bibnamefont{Saunders}},
  \bibinfo{journal}{Phys. Rev. Lett.} \textbf{\bibinfo{volume}{90}},
  \bibinfo{pages}{115301} (\bibinfo{year}{2003}).

\bibitem[{\citenamefont{Friman and Krotscheck}(1982)}]{Bengt}
\bibinfo{author}{\bibfnamefont{B.~L.} \bibnamefont{Friman}} \bibnamefont{and}
  \bibinfo{author}{\bibfnamefont{E.}~\bibnamefont{Krotscheck}},
  \bibinfo{journal}{Phys. Rev. Lett.} \textbf{\bibinfo{volume}{49}},
  \bibinfo{pages}{1705} (\bibinfo{year}{1982}).

\bibitem[{\citenamefont{Krotscheck and Springer}(2003)}]{he3mass}
\bibinfo{author}{\bibfnamefont{E.}~\bibnamefont{Krotscheck}} \bibnamefont{and}
  \bibinfo{author}{\bibfnamefont{J.}~\bibnamefont{Springer}},
  \bibinfo{journal}{J. Low Temp. Phys.}  (\bibinfo{year}{2003}),
  \bibinfo{note}{in press}.

\bibitem[{\citenamefont{Fetter and Walecka}(1971)}]{FetterWalecka}
\bibinfo{author}{\bibfnamefont{A.~L.} \bibnamefont{Fetter}} \bibnamefont{and}
  \bibinfo{author}{\bibfnamefont{J.~D.} \bibnamefont{Walecka}},
  \emph{\bibinfo{title}{Quantum Theory of Many-Particle Systems}}
  (\bibinfo{publisher}{McGraw-Hill}, \bibinfo{address}{New York},
  \bibinfo{year}{1971}).

\bibitem[{\citenamefont{Apaja et~al.}(1997)\citenamefont{Apaja, Halinen,
  Halonen, Krotscheck, and Saarela}}]{bosegas}
\bibinfo{author}{\bibfnamefont{V.}~\bibnamefont{Apaja}},
  \bibinfo{author}{\bibfnamefont{J.}~\bibnamefont{Halinen}},
  \bibinfo{author}{\bibfnamefont{V.}~\bibnamefont{Halonen}},
  \bibinfo{author}{\bibfnamefont{E.}~\bibnamefont{Krotscheck}},
  \bibnamefont{and} \bibinfo{author}{\bibfnamefont{M.}~\bibnamefont{Saarela}},
  \bibinfo{journal}{Phys. Rev. B} \textbf{\bibinfo{volume}{55}},
  \bibinfo{pages}{12925} (\bibinfo{year}{1997}).

\bibitem[{\citenamefont{Aldrich and Pines}(1976)}]{Aldrich}
\bibinfo{author}{\bibfnamefont{C.~H.} \bibnamefont{Aldrich}} \bibnamefont{and}
  \bibinfo{author}{\bibfnamefont{D.}~\bibnamefont{Pines}}, \bibinfo{journal}{J.
  Low Temp. Phys.} \textbf{\bibinfo{volume}{25}}, \bibinfo{pages}{677}
  (\bibinfo{year}{1976}).

\bibitem[{\citenamefont{Friman and Blaizot}(1981)}]{FrimanBlaizot}
\bibinfo{author}{\bibfnamefont{B.~L.} \bibnamefont{Friman}} \bibnamefont{and}
  \bibinfo{author}{\bibfnamefont{J.~P.} \bibnamefont{Blaizot}},
  \bibinfo{journal}{Nucl. Phys. A} \textbf{\bibinfo{volume}{372}},
  \bibinfo{pages}{69} (\bibinfo{year}{1981}).

\bibitem[{\citenamefont{Krotscheck et~al.}(1998)\citenamefont{Krotscheck,
  Paaso, Saarela, Sch{\"o}rkhuber, and Zillich}}]{mixmass}
\bibinfo{author}{\bibfnamefont{E.}~\bibnamefont{Krotscheck}},
  \bibinfo{author}{\bibfnamefont{J.}~\bibnamefont{Paaso}},
  \bibinfo{author}{\bibfnamefont{M.}~\bibnamefont{Saarela}},
  \bibinfo{author}{\bibfnamefont{K.}~\bibnamefont{Sch{\"o}rkhuber}},
  \bibnamefont{and} \bibinfo{author}{\bibfnamefont{R.}~\bibnamefont{Zillich}},
  \bibinfo{journal}{Phys. Rev. B} \textbf{\bibinfo{volume}{58}},
  \bibinfo{pages}{12282} (\bibinfo{year}{1998}).

\bibitem[{\citenamefont{Krotscheck and Sch{\"o}orkhuber}(2003)}]{KarSLT23}
\bibinfo{author}{\bibfnamefont{E.}~\bibnamefont{Krotscheck}} \bibnamefont{and}
  \bibinfo{author}{\bibfnamefont{K.}~\bibnamefont{Sch{\"o}orkhuber}},
  \bibinfo{journal}{Physica B} \textbf{\bibinfo{volume}{in press}}
  (\bibinfo{year}{2003}).

\bibitem[{\citenamefont{Brown et~al.}(1982)\citenamefont{Brown, Pethick, and
  Zaringhalam}}]{ZaringhalamMass}
\bibinfo{author}{\bibfnamefont{G.~E.} \bibnamefont{Brown}},
  \bibinfo{author}{\bibfnamefont{C.~J.} \bibnamefont{Pethick}},
  \bibnamefont{and}
  \bibinfo{author}{\bibfnamefont{A.}~\bibnamefont{Zaringhalam}},
  \bibinfo{journal}{J. Low Temp. Phys.} \textbf{\bibinfo{volume}{48}},
  \bibinfo{pages}{349} (\bibinfo{year}{1982}).

\bibitem[{\citenamefont{Mishra et~al.}(1983)\citenamefont{Mishra, Brown, and
  Pethick}}]{PethickMass}
\bibinfo{author}{\bibfnamefont{V.~K.} \bibnamefont{Mishra}},
  \bibinfo{author}{\bibfnamefont{G.~E.} \bibnamefont{Brown}}, \bibnamefont{and}
  \bibinfo{author}{\bibfnamefont{C.~J.} \bibnamefont{Pethick}},
  \bibinfo{journal}{J. Low Temp. Phys.} \textbf{\bibinfo{volume}{52}},
  \bibinfo{pages}{379} (\bibinfo{year}{1983}).

\bibitem[{\citenamefont{Greywall}(1990)}]{Greywall90}
\bibinfo{author}{\bibfnamefont{D.~S.} \bibnamefont{Greywall}},
  \bibinfo{journal}{Phys. Rev. B} \textbf{\bibinfo{volume}{41}},
  \bibinfo{pages}{1842} (\bibinfo{year}{1990}).

\bibitem[{\citenamefont{Morhard et~al.}(1996)\citenamefont{Morhard,
  B{\"a}uerle, Bossy, Bunkov, Fisher, and Godfrin}}]{Godfrin2dmass}
\bibinfo{author}{\bibfnamefont{K.-D.} \bibnamefont{Morhard}},
  \bibinfo{author}{\bibfnamefont{C.}~\bibnamefont{B{\"a}uerle}},
  \bibinfo{author}{\bibfnamefont{J.}~\bibnamefont{Bossy}},
  \bibinfo{author}{\bibfnamefont{Y.}~\bibnamefont{Bunkov}},
  \bibinfo{author}{\bibfnamefont{S.~N.} \bibnamefont{Fisher}},
  \bibnamefont{and} \bibinfo{author}{\bibfnamefont{H.}~\bibnamefont{Godfrin}},
  \bibinfo{journal}{Phys. Rev. B} \textbf{\bibinfo{volume}{53}},
  \bibinfo{pages}{2658} (\bibinfo{year}{1996}).

\bibitem[{\citenamefont{Greywall}(1983)}]{GRE83}
\bibinfo{author}{\bibfnamefont{D.~S.} \bibnamefont{Greywall}},
  \bibinfo{journal}{Phys. Rev. B} \textbf{\bibinfo{volume}{27}},
  \bibinfo{pages}{2747} (\bibinfo{year}{1983}).

\bibitem[{\citenamefont{Greywall}(1986)}]{GRE86}
\bibinfo{author}{\bibfnamefont{D.}~\bibnamefont{Greywall}},
  \bibinfo{journal}{Phys. Rev. B} \textbf{\bibinfo{volume}{33}},
  \bibinfo{pages}{7520} (\bibinfo{year}{1986}).

\bibitem[{\citenamefont{B{\"a}uerle et~al.}(1998)\citenamefont{B{\"a}uerle,
  Bunkov, Chen, Fisher, and Godfrin}}]{GodfrinULT98}
\bibinfo{author}{\bibfnamefont{C.}~\bibnamefont{B{\"a}uerle}},
  \bibinfo{author}{\bibfnamefont{Y.~M.} \bibnamefont{Bunkov}},
  \bibinfo{author}{\bibfnamefont{A.~S.} \bibnamefont{Chen}},
  \bibinfo{author}{\bibfnamefont{S.~N.} \bibnamefont{Fisher}},
  \bibnamefont{and} \bibinfo{author}{\bibfnamefont{H.}~\bibnamefont{Godfrin}},
  \bibinfo{journal}{J. Low Temp. Phys.} \textbf{\bibinfo{volume}{110}},
  \bibinfo{pages}{333} (\bibinfo{year}{1998}).

\end{thebibliography}

\end{document}